\documentstyle[12pt]{article}

\def\ar{\rightarrow}
\def\bib{\bibitem}

\def\intx{\int\! d^{\sl 4}x}
\def\intX{\int\! d^{\sl 4}X\,}

\def\pa{\partial}

\def\al{\alpha}
\def\be{\beta}
\def\ga{\gamma}
\def\de{\delta}

\def\la{\lambda}
\def\va{\varphi}

\def\Ga{{\it\Gamma}}
\def\La{{\it\Lambda}}

\def\beq{\begin{equation}}
\def\eeq{\end{equation}}
\def\bed{\begin{displaymath}}
\def\eed{\end{displaymath}}
\def\beqq{\begin{eqnarray}}
\def\eeqq{\end{eqnarray}}
\def\bedd{\begin{eqnarray*}}
\def\eedd{\end{eqnarray*}}

\begin{document}

\centerline{\normalsize\bf GENERAL RELATIVITY AS THE CLASSICAL LIMIT} \centerline{\normalsize\bf OF THE RENORMALIZABLE GAUGE THEORY} \centerline{\normalsize\bf OF VOLUME PRESERVING DIFFEOMORPHISMS}

\vspace*{0.9cm}
\centerline{\footnotesize C. WIESENDANGER}
\baselineskip=12pt
\centerline{\footnotesize\it Aurorastr. 24, CH-8032 Zurich}
\centerline{\footnotesize E-mail: christian.wiesendanger@ubs.com}

\vspace*{0.9cm}
\baselineskip=13pt
\abstract{The different roles and natures of spacetime appearing in a quantum field theory and in classical physics are analyzed implying that a quantum theory of gravitation is not necessarily a quantum theory of curved spacetime. Developing an alternative approach to quantum gravity starts with the postulate that inertial and gravitational energy-momentum need not be the same for virtual quantum states. Separating their roles naturally leads to the quantum gauge field theory of volume-preserving diffeomorphisms of an inner four-dimensional space. The classical limit of this theory coupled to a quantized scalar field is derived for an on-shell particle where inertial and gravitational energy-momentum coincide. In that process the symmetry under volume-preserving diffeomorphisms disappears and a new symmetry group emerges: the group of coordinate transformations of four-dimensional spacetime and with it General Relativity coupled to a classical relativistic point particle.}

\normalsize\baselineskip=15pt

\section{Introduction}
Spacetime is a basic ingredient in the construction of any quantum field theory (QFT) of microscopic interactions such as the electro-magnetic, weak and strong forces in the Standard Model (SM). This is immediately obvious looking at both the canonical or the path integral quantization approaches full of mathematical expressions such as Lagrangian densities, Fourier transforms, time-ordered products of quantum field operators etc which are all defined on a four-dimensional spacetime conventionally taken as Minkowski space \cite{stw1,stw2,cli,tpc}. What that spacetime - clearly not the same as macroscopic observable spacetime - , however, really is and how its geometrical and other properties can be established is normally not further reflected.

Looking at the experimental information we have about microscopic interactions of elementary particles which originates from scattering experiments what we observe is a number of incoming particles - typically two - characterized by their masses, four-momenta, electric charges etc transitioning with some probability into a number of outgoing particles again characterized by their masses, four-momenta, electric charges etc. The nature and properties of these particles are defined for the incoming ones by the accelerator beam characteristics, and are deduced for the outgoing ones from analyzing their observable macroscopic spacetime trajectories after the scattering event. What is not observable is the detailed spacetime evolution of the transition process which happens at spacetime distances escaping any experimental access and which remains a black box to the observer.

We note that the three-dimensional space in which the observed trajectories of the outgoing particles appear is the macroscopic space we abstract as part of an observable four-dimensional Minkowski space whose geometric properties we can establish experimentally. This is consistent with describing the trajectories as originating from classical relativistic point particles.

To establish a model for what happens in the unobservable black box one links the experimental information to the machinery of an appropriate QFT and its $S$-matrix by abstracting the incoming and outgoing particles as non-interacting asymptotic quantum states and employing the LSZ-reduction formalism to express the scattering amplitudes as Fourier-transformed, amputed, on-shell vacuum expectation values of time-ordered products of quantum field operators \cite{stw1,stw2,cli,tpc}. The asymptotic states have the characteristics such as mass, four-momentum, electric charge etc experimentally established as discussed above and live in an appropriate Fock space. If such a model - the prototype of which is the SM - is correct it allows us to predict the various observed transition probabilities.

The key point to note is that the spacetime necessarily appearing in the definition of the QFT giving us a model of the black box is a mathematical idealization and is {\it not} the same as the macroscopic Minkowski space in which the classical trajectories of the incoming and outgoing particles appear. Whether or not the idealization of the scattering event itself happening in an unobservably small spacetime region by means of a QFT defined on a spacetime idealized as another Minkowski space is correct can only be established after the facts, i.e. by the correctness of the predictions of the model.

Turning to gravitation spacetime is even more intimately woven into the construction of a field theory of gravity at the macroscopic level. Here the Equivalence Principle forces all classical particles to move on geodesics and implies the geometrization of gravity which is beautifully embodied in General Relativity (GR) \cite{stw3,lali,cmw}. Spacetime carries gravity by means of its non-Euclidean geometric structure which in turn is determined by the energy-momentum content of all non-gravitational matter and fields in spacetime.

We now look at the question of how gravity might microscopically interact with elementary particles and the other three microscopic interactions. If it remains true that we essentially want to describe events such as the scattering of elementary particles and the quanta of the gravitational interaction (as well as the quanta of the other three microscopic interactions) by gravity then a picture similar to the above should apply. So the experimental observation of macroscopic trajectories of scattered elementary particles should allow to assign properties to the asymptotic quantum states of an appropriate QFT model for the black box of the scattering event itself. Again spacetime with some geometric structure will enter that QFT model. But whether the idealization of the scattering event itself happening in an unobservably small spacetime region by means of a QFT defined on that spacetime will be correct can again only be established after the facts, i.e. by the correctness of the predictions of the model. As a result the structure of macroscopic spacetime has no a priori implication for the microscopic idealization of the spacetime entering such a QFT model.

So, a quantum theory of gravitation is {\it not} necessarily a quantum theory of curved spacetime. Instead any approach should be worthwhile to develop which respects the various conditions for a viable QFT such as causality, renormalizability and the validity of conservation laws such as for energy-momentum and which yields a classical limit respecting the Equivalence Principle, hence geometrizing gravity at the classical level. One such approach based on Minkowski space as the idealized spacetime embedded in the QFT model for gravity indeed exists. It takes as its basic postulate that the Equivalence Principle for observable physical states is necessarily valid, but for virtual quantum states it is not \cite{chw1,chw2,chw4}.

\section{Why should Inertial and Gravitational Mass be the Same for Virtual Quantum States?}

GR has been developed starting from the observed equality of inertial and gravitational mass $ m_I = m_G $ \cite{stw3,lali,cmw}. To be in agreement with observation this equality has to hold in any expression describing observable states in a gravitational context in their rest frames. However, in formulating a theory nothing enforces this equality for virtual (=non-observable) quantum states as long as it continues to hold for the on-shell (=observable) quantum states in that theory.

Now (a) the observed equality of inertial and gravitational mass of an on-shell physical object in its rest frame together with (b) the conservation of the inertial energy-momentum $ p_I^\mu $ in any frame tells us that in the rest frame
\beq \label{1}
p_I^\mu = (m_I,\underline{0}) =^{\!\!\!\!\!\!\! ^{ (a)}} (m_G,\underline{0}) = p_G^\mu
\eeq
assuming that the gravitational energy-momentum $ p_G^\mu $ plays a physical role different from that of the inertial energy-momentum, yet being observationally identical for on-shell objects. However, for off-shell states why shouldn't there be two separate conservation laws, one for the inertial energy-momentum and the other for the gravitational energy-momentum?

To explore this route let us postulate both $p_I^\mu$ and $p_G^\mu$ to be two separate four-vectors which are conserved, but in our approach through two different mechanisms. The conservation of $p_I^\mu$ is related to translation invariance in spacetime. Making use of Noether's theorem a second conserved four-vector can be constructed which is related to the invariance under volume-preserving diffeomorphisms of a four-dimensional inner space. That four-vector is then interpreted as the gravitational energy-momentum $p_G^\mu$ in the construction of a gauge theory of gravitation which we will review in section 4 below.

In a series of papers we have established this theory as the gauge theory of the group of volume-preserving diffeomorphisms at the classical \cite{chw1} and quantum level \cite{chw2} where we also have calculated the beta function to one loop which shows that the pure gauge field theory is asymptotically free whereas the theory including all SM fields is not. To proof mathematical consistency and to ensure prediction power for physical quantities in terms of the original couplings, masses etc. we then have demonstrated the renormalizability of the theory to all orders in perturbation theory \cite{chw4}. In these papers we have shown that one can consistently deal with the complications arising from a non-compact gauge group, e.g. ensuring the positivity of the gauge field Hamiltonian or regularizing divergent integrals over inner degrees of freedom related to the infinite volume of the gauge group which arise in a perturbative expansion.

Finally, the observed equality of inertial and gravitational energy-momentum in this approach is assured by introducing a physical limit for on-shell physical objects, the construction of which is based on the definition of observable asymptotic states and a suitable $S$-matrix \cite{chw3} which is shown there to be unitary. In essence the limit amounts to equaling inertial and gravitational energy-momentum, hence ensuring the validity of the Equivalence Principle.

Now does all of this really yield a quantum theory of gravity?

To properly answer this question we analyze below the classical limit $ \hbar \ar 0$ of the gauge theory of volume-preserving diffeomorphisms coupled to a scalar matter field. In section 3 we review the steps involved in taking the classical limit in scalar QED to re-iterate them in section 5 in the present case. There we will find crucial differences to the QED case which result in the disappearance of the symmetry of the theory under volume-preserving diffeomorphisms of an inner space we have started with. Instead a new symmetry group will emerge: the group of coordinate transformations of four-dimensional spacetime and with it General Relativity. Hence, as is necessary for the interpretation of the gauge theory of volume-preserving diffeomorphisms as a quantum theory of gravity GR emerges as its classical limit.

One final word on presentational style. Below we will use the same symbols for all: quantum field operators, classical fields and quantum mechanical (pseudo-)probability amplitudes. It will always be clear what is meant in which expression. However, for the sake of presentational clarity we have omitted the admittedly challenging "details" of normal ordering, gauge-fixing, employing Dirac brackets when quantizing etc so as to be able to focus on the main arguments. It is also understood that the physical limit has to be taken whenever aiming for the calculation of observable quantities.

\section{Classical Limit of a Scalar Quantum Field Coupled to Quantum Electrodynamics}
In this section we review the way from a theory given in terms of a quantized scalar field coupled to the quantized electromagnetic field back to a theory in terms of a classical point particle coupled to classical electrodynamics \cite{stw1,lali}.

To do so we start with the action for a charged scalar field $\phi (x)$ and its conjugate $\phi^\dagger (x)$ 
\beq \label{2}
S_M = - \intx\, \Big\{ \big[ i D_\mu (x) \phi (x) \big]^\dagger 
i D^\mu (x) \phi (x) +\, m^2 \, \phi^\dagger (x) \phi (x) \Big\}
\eeq
coupled to the electromagnetic field $A_\mu (x)$ with action
\beq \label{3}
S_G = -\frac{1}{4\, e^2} \intx\, F_{\mu\nu} (x) \, F^{\mu\nu} (x).
\eeq
Both fields are defined on an idealized unobservable four-dimensional Minkowski spacetime ${\bf M^{\sl 4}}$ as discussed in the introduction. And both $\phi (x)$ and $A_\mu (x)$ are non-commuting quantum field operators acting on a suitable Fock space and subject to canonical commutation relations - hence, the expressions above are to be taken with a grain of salt as stated in the introduction.

Above
\beq \label{4}
D_\mu (x) = \pa_\mu + i A_\mu (x) 
\eeq
denotes the covariant derivative and
\beq \label{5}
F_{\mu\nu} (x) = \pa_\mu A_\nu (x) - \pa_\nu A_\mu (x)
\eeq
the field strength operator.

The action $S_M + S_G$ is by inspection invariant under the combined gauge transformations
\beqq \label{6}
A_\mu (x) &\ar& A_\mu (x) + \pa_\mu \La (x) \\
\phi (x) &\ar& e^{-i \La (x)}\, \phi (x) \nonumber
\eeqq
with $\La (x)$ a suitable scalar function.

Though somewhat formal we can write down the field equations for the quantum field operators $\phi (x)$
\beq \label{7}
-\, (i \pa_\mu - A_\mu (x) )\, (i \pa^\mu - A^\mu (x) )\, \phi (x) 
-\, m^2\, \phi (x) = 0
\eeq
and $A_\mu (x)$
\beq \label{8}
\frac{1}{e^2}\, \pa^\mu F_{\mu\nu} (x) = j_\nu (x),
\quad j_\nu (x) = -\frac{\de S_M}{\de A^\nu (x)} 
\eeq
which follow from varying the action $S_M + S_G$. It is the field equation for $\phi (x)$ from which one recovers the relativistic Hamiltonian for a classical charged point particle \cite{stw1,lali}.

Let us turn to the classical limit $\hbar \ar 0$ which can be thought to come about in two steps with profound implications on the observability of spacetime.

In the first step "second quantization" is reversed and the non-commu- ting field operators $\phi (x)$ and $A_\mu (x)$ which are subject to canonical commutation relations become commuting fields. $\phi (x)$ becomes a (pseudo-) probability amplitude for a charged relativistic quantum-mechanical point particle and $A_\mu (x)$ becomes a classical Maxwell field. Both commuting fields are again described by the actions Eqn.(\ref{2}) and Eqn.(\ref{3}) which are invariant under the gauge transformations Eqns.(\ref{6}) and by the corresponding field equations Eqn.(\ref{7}) and Eqn.(\ref{8}) - however, mathematically and physically their interpretation is now a very different one.

In the second step "first quantization" of the point particle is reversed as well and its non-commuting position and momentum operators $x_\mu$ and $p^\nu$ subject to the canonical commutation relation $[x_\mu,p^\nu] = -i\, \eta_\mu\,\!^\nu$ become commuting c-numbers which is reflected by the replacement
\beq \label{9}
i \pa^\mu \leftrightarrow p^\mu
\eeq
based on the Correspondence Principle. $x_\mu$ simply becomes the position, $p^\nu$ the momentum of a charged relativistic point particle. $A_\mu (x)$ remains the classical Maxwell field. 

After the last step the field equation Eqn.(\ref{7}) for the probability amplitude for a single charged relativistic quantum-mechanical particle is transformed into the relativistic Hamiltonian for a classical charged point particle 
\beq \label{10}
\frac{1}{2\, m}\, (p_\mu - A_\mu (x) )\, (p^\mu - A^\mu (x) ) 
+ \frac{1}{2}\, m = 0
\eeq
after multiplying with $\frac{1}{2\, m}$. This equation is now defined on macroscopic observable Minkowski spacetime.

Let us next assume Eqn.(\ref{10}) is all we know about the motion of a classical charged point particle in the background of a vector field $A_\mu (x)$. We then see that the corresponding Hamiltonian equation of motion derived from Eqn.(\ref{10})
\beq \label{11}
m\, \ddot x_\mu = F_{\mu\nu} (x)\, \dot x^\nu
\eeq
with $F_{\mu\nu}$ as in Eqn.(\ref{5}) is invariant under a gauge transformation
\beq \label{12}
A_\mu (x) \ar A_\mu (x) + \pa_\mu \La (x)
\eeq
of the vector field. Now it cannot be that the trajectory of the point particle is not dependent on gauge transformations whereas the dynamics of the vector field is. So it is natural to look for vector field actions invariant under Eqn.(\ref{12}). The action of lowest mass dimension is then immediately found to be \cite{stw1,lali}
\beq \label{13}
S_G \propto -\intx\, F_{\mu\nu} (x) \, F^{\mu\nu} (x) + O (F^3)
\eeq
with the higher order terms suppressed at sufficiently low energies. As $\intx\, F_{\mu\nu} (x) \, F^{\mu\nu} (x)$ is dimensionless, any coupling constant in the leading order term has to be dimensionless too and the reasoning above allows one to immediately recover the gauge field action Eqn.(\ref{3}).

The case of a quantized scalar field coupled to the quantized electromagnetic field does not generate new insights by itself, but for the clarification of the different natures of spacetime entering the quantum field theoretical and the classical descriptions. And it paves the way for a similar reasoning in the case of a quantized scalar field coupled to the quantized gauge field for the gauge theory of volume-preserving diffeomorphisms. There new things will happen which result in GR emerging in the classical limit.

\section{Gauge Theory of Volume-Preserving Diffeomorphisms Revisited}
In this section we revisit the basics of the gauge theory of volume-preserving diffeomorphisms developed in \cite{chw1,chw2,chw4,chw3}.

As discussed in section 2 we want to explore what happens when keeping inertial and gravitational energy-momentum as separate entities in a physical theory - taking the physical limit for observable quantities to ensure their equality as demanded by the Equivalence Principle. As both types of momentum are conserved we first need to establish two separate conservation laws for two four-vectors. Obviously conservation of inertial energy-momentum is related by Noether's theorem to global spacetime translation invariance of the theory. Employing Noether's theorem a second time to generate another conserved four-vector requires invariance of the theory under an independent second translation group. To be specific we take this group to be the larger group of volume-preserving diffeomorphisms of a four-dimensional space $V \subset {\bf R^{\sl 4}}$ whose coordinates are labelled by $X^\al$. As to further notations we refer to the Appendix below.

Infinitesimal group transformations can be written as 
\beq \label{14}
X^\al\ar X^\al + {\cal E}^\al (X),\:\:\al=\sl{0,1,2,3}
\eeq
where the condition on the infinitesimal translation parameter ${\cal E}^\al (X)$
\beq \label{15}
\nabla_\al {\cal E}^\al (X) = 0
\eeq
ensures volume preservation. 

To represent this group on the various fields in the theory we need to add the necessary inner degrees of freedom so that all fields $\phi (x,X)$ are defined on the product of an idealized unobservable four-dimensional Minkowski spacetime ${\bf M^{\sl 4}}$ times the additional inner four-dimensional space $V\subset {\bf R^{\sl 4}}$ we require to have finite volume. The $X^\al$ in $\phi (x,X) \sim \phi_X (x)$ labelling continous vectors can be thought of as a generalization of the inner indices $a$ in $\phi_a (x)$ labelling discrete vectors in the context of a Yang-Mills theory.

The action for a field $\phi (x,X)$
\beq \label{16}
S_M = \intx \intX \La^{-4}\, {\cal L}_M (\phi (x,X), \pa_\mu \phi (x,X)) \nonumber
\eeq
is then automatically invariant under both global spacetime translations and volume-preserving diffeomorphisms which act on the field $\phi (x,X)$ as
\beqq \label{17}
x^\mu &\ar& x'^\mu = x^\mu, \quad
X^\al \ar X'^\al = X^\al, \\
\phi (x,X) &\ar& \phi (x,X) -\, {\cal E}^\al (X)\cdot \nabla_\al\, \phi (x,X). \nonumber
\eeqq
These invariances generate the independent conservation laws for two four-vectors \cite{chw1}.

Above $\La$ is a parameter with dimension of length introduced to keep the volume element in inner space dimensionless. The volume integration $\intX \La^{-4}$ or sum over the continous indices $X^\al$ is nothing but the generalization of a sum over inner indices $\sum_a$ in Yang-Mills theories. In \cite{chw1} we have shown the theory to be scale-invariant in inner space when rescaling $\La$ at the same time as inner coordinates and fields so that $\La$ can be chosen arbitrarily. Note that the finite volume of the inner space $V$ can always taken to be equal to $\La^4$.

Next we take the infinitesimal translations or gauge parameters ${\cal E}^\al (X)$ local
\beq \label{18}
{\cal E}^\al (X) \ar {\cal E}^\al (x,X).
\eeq 
${\cal E}^\al (x,X)$ still obeys Eqn.(\ref{15}) and Eqns.(\ref{17}) still define the group representation on the fields.

This requires the introduction of a covariant derivative
\beq \label{19}
D_\mu (x,X) = \pa_\mu + A_\mu\,^\al (x,X) \nabla_\al 
\eeq
to preserve the invariance of the action Eqn.(\ref{16}) under local group or gauge transformations which is achieved by replacing $\pa_\mu\ar D_\mu (x,X)$ so that we now have
\beq \label{20}
S_M = \intx\intX \La^{-4}\, {\cal L}_M (\phi (x,X), D_\mu (x,X) \phi (x,X)). \nonumber
\eeq

Above $ \nabla_\al = \frac{\pa\,\,\,}{\pa X^\al} $ denote the generators of the inner translations. The gauge fields $A_\mu\,^\al (x,X)$ introduced in the process have to transform under local gauge transformations as
\beqq \label{21}
\!\!\!\!\!\! A_\mu\,^\al (x,X) \!\! &\ar& \!\! A_\mu\,^\al (x,X) + \pa_\mu {\cal E}^\al (x,X) + A_\mu\,^\be (x,X) \cdot \nabla_\be {\cal E}^\al (x,X) 
\nonumber \\
& &\quad\quad\quad -\, {\cal E}^\be (x,X) \cdot \nabla_\be A_\mu\,^\al (x,X)
\eeqq
and obey the divergence-free condition
\beq \label{22}
\nabla_\al A_\mu\,^\al (x,X) = 0
\eeq
as do all fields living in the gauge algebra.

The homogenously transforming field strength components are then found to be
\beqq \label{23}
& &\quad\quad F_{\mu\nu}\,^\al (x,X)= \pa_\mu A_\nu\,^\al (x,X) 
- \pa_\nu A_\mu\,^\al (x,X) \\
& & + A_\mu\,^\be (x,X)\cdot \nabla_\be A_\nu\,^\al (x,X)
- A_\nu\,^\be (x,X)\cdot \nabla_\be A_\mu\,^\al(x,X) \nonumber 
\eeqq
in terms of which the gauge field action of lowest mass dimension is \cite{chw1}
\beq \label{24}
S_G = -\frac{1}{4\, \La^2} \intx\intX \La^{-4}\, F_{\mu\nu}\,^\al (x,X) \cdot F^{\mu\nu}\,_\al (x,X).
\eeq
To be precise the above $S_G$ is given in terms of an inner metric $\eta$ which is used to raise and lower inner indices after a partial gauge-fixing to the so-called Minkowski gauges which preserve $\eta$ \cite{chw1}.

In \cite{chw4} we have shown that theories defined by $S_M + S_G$ are renormalizable as long as $S_M$ contains fields $\phi (x,X)$ and their spacetime derivatives of mass dimension four or less only. Note that to obtain physical observables and to implement the Equivalence Principle those physical observables obey we have to take the physical limit ensuring equality of inertial and gravitational energy-momentum at the end of all calculations. Its exact meaning in the process of defining a unitary $S$-matrix for quantum gravity has been established in \cite{chw3}.

\section{Classical Limit of a Scalar Quantum Field Coupled to the Quantum Gauge Field Theory of Volume Preserving Diffeomorphisms}
In this section we derive the classical limit of a theory given in terms of a quantized scalar field coupled to the quantized gauge fields of the gauge theory of volume-preserving diffeomorphisms. In that process the inner space collapses, the field dependence on inner coordinates disappears and so does the symmetry under volume-preserving diffeomorphisms of the inner space. On the other hand a new symmetry group emerges: the group of coordinate transformations of four-dimensional spacetime and with it General Relativity coupled to a classical relativistic point particle.

In order to follow $\hbar\ar 0$ when deriving the classical limit in all expressions we reinstall the factors of $\hbar$ and take along factors of $c$, the speed of light, and $\Ga$, the gravitational constant.

Let us start again with the action for a scalar field $\phi (x,X)$ and its conjugate $\phi^\dagger (x,X)$
\beqq \label{25}
& &\!\!\!\!\!\!\!\!\!\!\!\!
S_M = - \intx\intX \La_P^{-4}\, \Big\{ c^2 \big[ i\hbar D_\mu (x,X) 
\phi (x,X)\big]^\dagger i\hbar D^\mu (x,X) \phi (x,X) \nonumber \\
& & \quad\quad\quad\quad\quad\quad\quad\quad +\, m^2 c^4 \, \va^\dagger (x,X) \va (x,X) \Big\} 
\eeqq
coupled to the gauge field $A_\mu\,^\al (x,X)$ with action
\beq \label{26}
S_G = -\frac{\hbar}{4\, g^2 \La_P^2} \intx\intX \La_P^{-4}\, F_{\mu\nu}\,^\al (x,X) \cdot F^{\mu\nu}\,_\al (x,X).
\eeq
Above we have taken $\La$ to equal the Planck length $\La_P$
\beq \label{27}
\La_P = \sqrt{\frac{\hbar \Ga}{c^3}}
\eeq
based on the inner scale invariance of the theory, have inserted factors of $\hbar$ to get the dimensions right and introduced a dimensionless coupling $g^2$. Note that
\beq \label{28}
\frac{\hbar}{4\, g^2 \La_P^2} = \frac{c^3}{16\, \pi \Ga}
\eeq
for the choice $g^2 = 4\, \pi$ ensuring the correct Newtonian limit as demonstrated in \cite{chw5}. Note in addition that this expression is independent of $\hbar$.

Above both fields $\phi (x,X)$ and $A_\mu\,^\al (x,X)$ are non-commuting quantum field operators defined on an idealized unobservable four-dimensional Minkowski spacetime times an inner space, acting on a suitable Fock space and subject to canonical commutation relations - hence, as in the case of electrodynamics the expressions above are to be taken with a grain of salt and have been properly elaborated in \cite{chw3}.

Though somewhat formal we can next write down the field equations for the quantum field operators $\phi (x,X)$
\beqq \label{29}
& & -\, c^2\, i\hbar (\pa_\mu + A_\mu\,^\al (x,X) \nabla_\al)\,
i\hbar (\pa^\mu + A^{\mu\,\be} (x,X) \nabla_\be)\, \phi (x,X) \nonumber \\
& & \quad\quad\quad\quad\quad\quad\quad\quad -\, m^2 c^4 \, \phi (x,X) = 0
\eeqq
and $A_\mu\,^\al (x,X)$
\beqq \label{30}
& & \frac{c^3}{4\, \pi \Ga}\, \left( {\cal D}^\mu (x,X) \right)^\al\,\!_\be  F_{\mu\nu}\,^\be (x,X) = j_\nu\,^\al (x,X), \\
& &\quad\quad\quad j_\nu\,^\al (x,X) = -\frac{\de S_M}{\de A^\nu\,_\al (x,X)}, \nonumber
\eeqq
where we have introduced the covariant derivative in the adjoint representation
\beq \label{31}
\left( {\cal D}_\mu (x,X) \right)^\al\,\!_\be = \left( \pa_\mu + A_\mu\,^\ga (x,X) \nabla_\ga \right) \eta^\al\,\!_\be - \nabla_\be A_\mu\,^\al (x,X).
\eeq

Let us turn to the classical limit $\hbar \ar 0$ which can be thought to come about in the same two steps as in the case of electrodynamics again with profound implications on the observability of spacetime, but this time with additional complications.

In the first step "second quantization" is reversed and the non-commu- ting field operators $\phi (x,X)$ and $A_\mu\,^\al (x,X)$ which are subject to canonical commutation relations become commuting fields. $\phi (x,X)$ becomes a (pseudo-)probability amplitude for a single relativistic quantum-mechani- cal particle in the background of a classical gauge field $A_\mu\,^\al (x,X)$. Both commuting fields are again described by the actions Eqn.(\ref{25}) and Eqn.(\ref{26}) which are invariant under the gauge transformations Eqns.(\ref{17}) and Eqn.(\ref{21}) and by the corresponding field equations Eqn.(\ref{29}) and Eqn.(\ref{30}) - however, mathematically and physically their interpretation in the physical limit is now a very different one.

In the second step "first quantization" of the point particle is reversed as well and its non-commuting position and momentum operators $x_\mu$ and $p^\nu$ subject to the canonical commutation relation $[x_\mu,p^\nu] = -i\hbar\, \eta_\mu\,\!^\nu$ as well as its non-commuting inner coordinate operators $X_\al$ and $P^\be$ subject to the canonical commutation relation $[X_\al,P^\be] = -i\hbar\, \eta_\al\,\!^\be$ become commuting c-numbers which is reflected by the replacements
\beq \label{32}
i\hbar \pa^\mu \leftrightarrow p^\mu,
\quad i\hbar \nabla^\al \leftrightarrow P^\al 
\eeq 
based on the Correspondence Principle. $x_\mu$ simply becomes the position, $p^\nu$ the momentum - with $X_\al$ and $P^\be$ their inner analogs - of a single relativistic classical particle. $A_\mu\,^\al (x,X)$ remains a classical gauge field. 

After multiplying with $\frac{1}{2\, m\, c^2}$ the field equation Eqn.(\ref{29}) for the probability amplitude for a single charged relativistic quantum-mechanical particle is transformed into the Hamiltonian for a relativistic classical point particle
\beq \label{33}
\frac{1}{2\, m}\, (p_\mu + A_{\mu\,\al} (x,X) P^\al)\,
(p^\mu + A^\mu\,_\be (x,X) P^\be) +\frac{1}{2}\, m\, c^2 = 0
\eeq
which now depends on the inner coordinates $X_\al$ and $P^\be$ as well.

So far nothing dramatic in comparison to the electrodynamics case has happened. However, we have not taken into account yet that (A) with $\hbar\ar 0$ the Planck length $\La_P = \sqrt{\frac{\hbar \Ga}{c^3}}\ar 0$ and with it the finite inner space collapses to a point and (B) that a classical point particle is by definition always observable so that we have to take the physical limit in the Hamiltonian above. Note that the argument does not depend on taking $\La = \La_P$ as any length $\La$ is proportional to $\La_P$ which tends to zero with $\hbar\ar 0$ taking the volume of the inner space $V$ to zero as well which is proportional to $\La_P^4\propto \hbar^2$.

The collaps of the inner space due to (A) implies that the field dependence on inner coordinates disappears: $A_\mu\,^\al (x,X)\ar A_\mu\,^\al (x)$, and with it the symmetry under volume-preserving diffeomorphisms of the inner space and (B) tells us to take the physical limit
\beq \label{34}
P^\al\ar p^\al.
\eeq
As a result Eqn.(\ref{33}) becomes
\beq \label{35}
\frac{1}{2\, m}\, (p_\mu + A_{\mu\,\al} (x) p^\al)\,
(p^\mu + A^\mu\,_\be (x) p^\be) +\frac{1}{2}\, m\, c^2 = 0.
\eeq
This equation is now defined on a macroscopic observable spacetime the geometry of which becomes evident below.

We can re-write this in a more perspicuous form in terms of
\beq \label{36}
e^\mu\,_\al (x) = \eta^\mu\,_\al + A^\mu\,_\al (x)
\eeq 
or of
\beq \label{37}
g_{\al\be} (x) = (\eta_{\mu\,\al} + A_{\mu\,\al} (x))\,
(\eta^\mu\,_\be + A^\mu\,_\be (x))
\eeq
as
\beqq \label{38}
& & \frac{1}{2\, m}\, p^\al\, \underbrace{ (\eta_{\mu\,\al} + A_{\mu\,\al} (x))\, (\eta^\mu\,_\be + A^\mu\,_\be (x))}\, p^\be +\frac{1}{2}\, m\, c^2 = 0 \nonumber \\
& &\quad\quad\quad\quad \frac{1}{2\, m}\, p^\al\, \underbrace{ e_{\mu\,\al} (x)\, e^\mu\,_\be (x)}\, p^\be +\frac{1}{2}\, m\, c^2 = 0 \\
& &\quad\quad\quad\quad\quad\quad\,\, \frac{1}{2\, m}\, p^\al\, g_{\al\be} (x)\, p^\be +\frac{1}{2}\, m\, c^2 = 0. \nonumber
\eeqq
As we will see $e^\mu\,_\al (x)$ has to be interpreted as a vierbein and $g_{\al\be} (x)$ as a metric on the observable macroscopic spacetime which leads us back to GR \cite{stw3,lali}.

To get there let us derive the Hamiltonian equations from the last expression above
\beqq \label{39}
\frac{\pa H}{\pa p^\ga} &=& \frac{1}{m}\, g_{\ga\be} (x)\, p^\be = \dot x_\ga \\
\label{40} \frac{\pa H}{\pa x^\ga} &=& \frac{1}{2\, m}\, p^\al p^\be\, \frac{g_{\al\be}(x)}{\pa x^\ga} = - \dot p_\ga.
\eeqq
The first equation tells us that if we interpret $g_{\ga\be} (x)$ as a metric, $p^\al$  and $\dot x_\ga$ transform under a general coordinate transformation
\beq \label{41}
x^\al \ar x'^\al (x),\quad g_{\al\be}(x) \ar g'_{\al\be}(x') = \frac{\pa x^\ga}{\pa x'^\al}\, \frac{\pa x^\de}{\pa x'^\be}\, g_{\ga\de}(x)
\eeq
as contra- and covariant vectors respectively
\beq \label{42}
p^\al \ar p'^\al = \frac{\pa x'^\al}{\pa x^\be}\, p^\be,\quad
\dot x_\al \ar \dot x'_\al = \frac{\pa x^\be}{\pa x'^\al}\, x_\be.
\eeq
In addition the Hamiltonian equations tell us that the point particle moves on geodesics of the metric $g_{\al\be} (x)$
\beq \label{43}
\ddot x^\ga = -\Ga^\ga_{\al\be} (x) \, \dot x^\al \dot x^\be,
\eeq
where
\beq \label{44}
\Ga^\ga_{\al\be} (x)
= \frac{1}{2}\, g^{\ga\de} \left\{ \frac{\pa g_{\de\al} }{\pa x^\be}
+ \frac{\pa g_{\de\be} }{\pa x^\al}  -\frac{\pa g_{\al\be} }{\pa x^\de}
\right\}
\eeq 
are the usual non-covariant Christoffel symbols.

So at the same time as the inner space has collapsed and the field dependence on inner coordinates has disappeared - and with it the symmetry under volume-preserving diffeomorphisms of the inner space - we see a new symmetry group emerging: the group of coordinate transformations of macroscopic observable four-dimensional spacetime and with it a classical relativistic point particle moving on geodesics as required by the Equivalence Principle. In fact it should better be like this, as the physical limit Eqn.(\ref{34}) we have taken to arrive at Eqn.(\ref{35}) expresses nothing but the equivalence of inertial and gravitational mass for observable particles which forces the latter to move on geodesics.

Now it cannot be that the trajectory of the point particle is not dependent on general coordinate transformations whereas the dynamics of the metric field $g_{\al\be} (x)$ is. So it is natural to look for actions for the metric invariant under Eqn.(\ref{41}). The action of lowest mass dimension is then immediately found to be \cite{stw1,lali}
\beq \label{45}
S_G \propto -\intx\, \sqrt{-g (x)}\, R (x) + O (R^2),
\eeq
where $R$ denotes the scalar curvature. The terms of higher order in the curvature tensor are suppressed at sufficiently low energies.

As $\intx\, \sqrt{-g (x)}\, R (x)$ carries the dimension of length to the power minus two, any coupling constant in the leading order term has to have dimension of energy times time and dimensional analysis allows one to infer that it has to be proportional to $\frac{c^3}{\Ga}$ recovering the Einstein-Hilbert action for GR.

We finally note that GR emerging in the classical limit does not depend on the order of taking the various limits in the reasoning above.

\section{Conclusions}
In this paper we have first clarified the different natures of spacetime entering a QFT model at the microscopic level versus entering the description of a relativistic point particle in the background of a classical field.

Taking the different natures of spacetime into account we then have established GR to emerge as the classical limit of the gauge field theory of volume-preserving diffeomorphisms coupled to a scalar field. To get there we have reiterated the way from scalar QED back to a classical relativistic point particle coupled to classical Electrodynamics. In that process two crucial differences to the QED case occur: on the one hand the inner space needed to represent the gauge group on fields collapses, the field dependence on inner coordinates disappears and so does the symmetry under volume-preserving diffeomorphisms of the inner space. On the other hand a new symmetry group emerges: the group of coordinate transformations of macroscopic observable four-dimensional spacetime and with it General Relativity coupled to a classical relativistic point particle. Note that the argument does not depend on the scalar nature of the matter field and that one gets the same result e.g. for spinors when $\hbar \ar 0$ as spin terms are of $O(\hbar)$.

It is reassuring that not only the microscopic strong and electro-weak interactions can be described within a renormalizable quantum gauge field theory framework formulated on an idealized unobservable Minkowski spacetime. In fact gravity at the quantum level can be described by following exactly the same logic, however, the theory gets more complicated due to its non-compact gauge group having an infinite volume. Yet it is still renormalizable. So Nature seems to allow for a consistent, rupture-free picture based on conservation laws and symmetry considerations at least up to energy scales far beyond experimental reach.

\appendix

\section{Notations and Conventions}

Generally, ({\bf M}$^{\sl 4}$,\,$\eta$) denotes the four-dimensional Minkowski space with metric $\eta=\mbox{diag}(-1,1,1,1)$, small letters denote space-time coordinates and parameters and capital letters denote coordinates and parameters in inner space.

Specifically, $x^\la,y^\mu,z^\nu,\dots\,$ denote Cartesian spacetime coordinates. The small Greek indices $\la,\mu,\nu,\dots$ from the middle of the Greek alphabet run over $\sl{0,1,2,3}$. They are raised and lowered with $\eta$, i.e. $x_\mu=\eta_{\mu\nu}\, x^\nu$ etc. and transform covariantly w.r.t. the Lorentz group $SO(\sl{1,3})$. Partial differentiation w.r.t to $x^\mu$ is denoted by $\pa_\mu \equiv \frac{\pa\,\,\,}{\pa x^\mu}$.

$X^\al, Y^\be, Z^\ga,\dots\,$ denote inner Cartesian coordinates we can always choose by partially fixing the gauge to so-called Minkowskian gauges \cite{chw1}. The small Greek indices $\al,\be,\ga,\dots$ from the beginning of the Greek alphabet run again over $\sl{0,1,2,3}$. They are raised and lowered with $\eta$, i.e. $x_\al=\eta_{\al\be}\, x^\be$ etc. and transform covariantly w.r.t. the inner Lorentz group $SO(\sl{1,3})$. Partial differentiation w.r.t to $X^\al$ is denoted by $\nabla_\al \equiv \frac{\pa\,\,\,}{\pa X^\al}$. 

The same lower and upper indices are summed unless indicated otherwise.

\end{document}